\documentclass{mem}
\usepackage{natbib}\usepackage{txfonts}\usepackage{balance}
\usepackage{graphicx,gensymb}
\usepackage[a4paper,breaklinks,dvipdfm]{hyperref}
\idline{83}{1}
\begin{document}
\def\teff{$T\rm_{eff }$}
\def\kms{$\mathrm {km s}^{-1}$}

\title{Radio--optical outlier quasars -- a case study with ICRF2 and SDSS}
	\subtitle{}
	
\author{
G.~Orosz\inst{1} 
\and S.~Frey\inst{2}
}

\offprints{S. Frey} 
 
\institute{
Department of Geodesy and Surveying, Budapest University of Technology and Economics, P.O. Box 91, H-1521 Budapest, Hungary
\and
F{\"O}MI Satellite Geodetic Observatory \& MTA Research Group for Physical Geodesy and Geodynamics, P.O. Box 585, H-1592 Budapest, Hungary
\email{frey@sgo.fomi.hu}
}

\authorrunning{G.~Orosz and S.~Frey}
\titlerunning{Radio--optical outlier quasars}

\abstract{
With Gaia, it will become possible to directly link the radio and optical reference frames using a large number of common objects. For the most accurate radio--optical link, it is important to know the level of spatial coincidence between the quasars' optical positions, and the radio positions determined by Very Long Baseline Interferometry (VLBI) observations. The ``outlier'' objects, for which the positions are significantly offset at the two different electromagnetic wavebands, may be of astrophysical interest as well. Here we present a case study to compare the radio positions of $\sim$800 quasars common in the second realization of the International Celestial Reference Frame (ICRF2) and in the Sloan Digital Sky Survey Data Release 7 (SDSS DR7) catalogue. Compared to the radio ICRF2, the SDSS provides two orders of magnitude less accurate astrometric data in the optical. However, its extensive sky coverage and faint magnitude limit allow us to directly relate the positions of a large sample of radio sources. This way we provide an independent check of the overall accuracy of the SDSS positions and confirm that the astrometric calibration of the latest Data Release 8 (DR8) is poorer than that of the DR7. We find over 20 sources for which the optical and radio brightness peaks are apparently not coincident at least at the 3$\sigma$ level of SDSS DR7 positional accuracy, and briefly discuss the possible causes, including dual active galactic nuclei.
}
\maketitle{}

\section{Introduction}

The second realization of the International Celestial Reference Frame (ICRF2) in the radio waveband was adopted by the International Astronomical Union (IAU) in 2009 \citep{fey1}. The ICRF2 catalogue contains the precise positions of 3414 compact extragalactic radio sources (active galactic nuclei, AGNs) observed with Very Long Baseline Interferometry (VLBI). At present, the most accurate optical astrometric catalogues cannot be linked directly (i.e. based on a set of overlapping objects) to the ICRF2 because the AGNs are too faint in the optical.
However, with the sensitive next-generation space-based optical astrometry mission, the European Space Agency's Gaia spacecraft \citep[e.g.][]{gaia} to be launched in 2013, it will become possible to directly link the radio and optical reference frames using a large number of common objects. This is important for astrometric reasons, to establish continuity between the radio and optical quasi-inertial reference frames, and for astrophysical applications, such as studying frequency-dependent AGN structure variations. 
While waiting for the Gaia extragalactic reference frame to be constructed around 2020, one can perform case studies for the reference frame link using presently available large optical sky surveys. On the one hand, these studies provide an independent assessment of the astrometric accuracy of the optical surveys. On the other hand, the comparison of the optical and radio positions of certain potential link sources may reveal significant non-coincidences. Finding the explanation(s) for such positional offsets prior to the availability of the Gaia reference frame will be an essential contribution for achieving the best radio--optical link in the future.

\section{A case study: directly linking ICRF2 and SDSS DR7}
\label{dr7}

We modelled the reference frame link using the 7$^{\rm th}$ Data Release (DR7) of the Sloan Digital Sky Survey (SDSS), which covers one-third ($\sim$12\,000 square degrees) of the sky in the optical, mainly in the Northern hemisphere and around the Equator \citep{abazajian}. The large sky coverage and the faint ($V$$\approx$$22^{\rm m}$) limiting magnitude make it possible to identify the counterparts of many radio-loud AGNs that have accurate radio positions available in the ICRF2 catalogue \citep{fey1}. The astrometric calibration of SDSS is described in detail by \citet{pier}. The source positions are calibrated using the 2$^{\rm nd}$ release of the USNO CCD Astrograph Catalogue \citep[UCAC2,][]{zacharias}, the UCAC r14 catalogue (a supplemental set of UCAC), and the SDSS+USNO-B catalogue \citep{munn}.

The typical formal positional accuracy of the individual radio sources in the ICRF2 is in the order of 100 microarcseconds ($\mu$as) in both right ascension ($\alpha$) and declination ($\delta$). For the SDSS DR7 in the optical, this value is two orders of magnitude higher, $\sim$50 milliarcseconds (mas). Among the all-sky set of 3414 ICRF2 sources, optical counterparts of 1014 were found in the SDSS DR7 within a search radius of $0\farcs5 = 500$~mas. 
The majority of our sources are in the range 7$^{\rm h}$$<$$\,\alpha\,$$<$18$^{\rm h}$ and $-5\degree$$<$$\,\delta\,$$<$$+70\degree$, with additional sources evenly distributed in right ascension at $-15\degree$$<$$\,\delta\,$$<$$+25\degree$ (this reflects the SDSS DR7 sky coverage). From the 1014 objects found, of which 208 are classified as extended (i.e. galaxies) and 806 as point-like (i.e. quasars) in the SDSS, only the latter, the unresolved sources are discussed here. These have somewhat more accurate optical coordinates. 

\begin{figure}[]
\resizebox{\hsize}{!}{\includegraphics[clip=true]{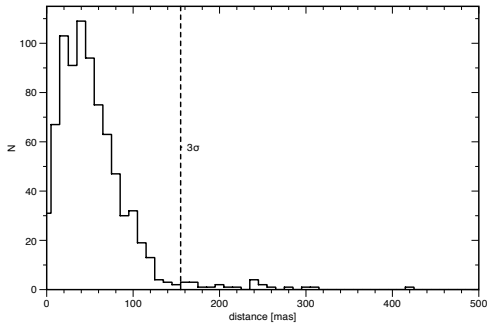}}
\caption{\footnotesize
Histogram of optical--VLBI radial positional differences for 806 ICRF2 sources identified in SDSS DR7 as quasars, within a search radius of 500~mas. The distribution is consistent with 52~mas (1$\sigma$) accuracy in both right ascension and declination in SDSS, assuming Rayleigh distribution. 
The vertical dashed line indicates the 3$\sigma$ value.
}
\label{dr7histo}
\end{figure}

The distribution of the SDSS--VLBI position differences for the quasars is shown in a histogram in Fig.~\ref{dr7histo}. It is generally consistent with $\sim$50~mas positional uncertainty in each coordinate, in a good agreement with the values determined for the SDSS. The optical minus radio right ascension differences ($\Delta\alpha\cos\delta$) have $\sigma_{\Delta\alpha}$$=$$48$~mas standard deviation and a negligible $\mu_{\Delta\alpha}$$=$$-8$~mas mean value. For the declination differences, $\sigma_{\Delta\delta}$$=$$54$~mas and $\mu_{\Delta\delta}$$=$$4$~mas. Earlier studies using SDSS DR4 data with 524 matching quasars and galaxies \citep{frey1}, and SDSS DR5 data with 735 matching objects \citep{frey2} led to similar results.

The tail of the distribution of coordinate differences in Fig.~\ref{dr7histo} is worth noting. There are significantly more sources with $>$3$\sigma$ positional offsets than expected statistically. Assuming that the differences in the two coordinates are independent normal random variables, the Rayleigh distribution would give $\sim$2 sources above the 3$\sigma$ level, as opposed to the detected 23 ``outliers'' (i.e. quasars with radio--optical separation exceeding 155~mas). 

As a first step to find an explanation for this, we investigated whether the outliers can be caused by chance coincidences between unrelated ICRF2 and SDSS objects seen at about the same direction. We constructed eight ``false'' radio source lists by simply shifting $\alpha$ and $\delta$ for all the ICRF2 sources by large arbitrary amounts, $\pm(1, 2, 3, 4)\degree$. We then tried to find SDSS optical counterparts within a $0\farcs5$ radius for these fake ``objects'' (over 27\,000 in total). The simulation showed that the probability of chance coincidences is 0.05\%, i.e. less than 1 false optical identification is expected within our $0\farcs5$ search radius. This suggests that essentially all of our outliers in Fig.~\ref{dr7histo} are real identifications. (So significant astrometric miscalibrations in the corresponding SDSS scans are very unlikely.)

When comparing directly the radio and optical positions, we naturally assume as a first approximation that the optical and radio emission peaks physically coincide. Because the activity of these distant AGNs is driven by their central supermassive black holes, and confined into their close vicinity, this assumption seems plausible in general. However, as we have seen, there are objects for which it is not necessarily true. For high-accuracy reference frame interconnections in the future, it is essential to understand this behaviour and to identify the outliers. Such objects should be avoided when the radio and optical reference frames are linked. 

There are a couple of physical phenomena which may casue some offset between the optical and radio emission peaks for AGNs. For example, the systematic difference between the radio (8~GHz) and optical positions due to the opacity effect \citep[known as ``core shift'', e.g.][]{kovalev,sokolovsky} can amount to $\sim$0.1~mas. However, this is far too small compared to the offsets we found. Another explanation might be that radio AGNs with extended VLBI structure show systematically higher positional non-coincidence ($\sim$8~mas) between their radio and optical centres than compact radio sources \citep{dasilvaneto,camargo}. But this effect is still too small to explain our offsets at 100-mas levels. Strong gravitational lensing could in principle cause the large offsets seen in Fig.~\ref{dr7histo}, in a configuration where the compact radio emission originates from the background (lensed) source, while the optical is dominated by a radio-quiet but optically bright lensing galaxy in the foreground. Recently \citet{finet} found that the lensing event probability is $\sim$0.59\%. Lensing is therefore insufficient to explain most the 23 outliers detected among $\sim$800 SDSS--ICRF2 quasars. 

Dual AGNs can also be viable candidates for producing the large systematic offsets. The projected linear size corresponding to 150--300~mas angular size is $\sim$1.2--2.4~kpc, at the redshift of $z$$\approx$1, using the standard cosmological model. These correspond to sub-galactic sizes. Merging galaxies with such separations do exist. Radio emission from e.g. double-peaked narrow [O\,III] emission-line quasar pairs from SDSS is detected in $\sim$100 cases (C. Li et al. 2012, in preparation), although these are generally weak radio sources, unlike the ICRF2 objects. Another problem is that only a small fraction of AGNs ($\sim$0.04\%--0.1\%) are predicted as duals with kpc-scale separations at $z$$\sim$0.5--1.2, and even less have detectable double-peaked narrow emission lines \citep{yu}. Even within our small sample of outliers there might be dual AGNs. If this could be verified, with the advent of Gaia, a much larger and astrometrically much more accurate AGN data set could be examined for the optical--VLBI position differences. It can provide us with a new observing tool to select binary AGN candidates.

\section{A note on SDSS DR8 astrometry}

The DR8 is currently the latest data release of the SDSS, covering $\sim$14\,500 square degrees of the sky \citep{aihara1}. 
We performed the same 500-mas radius search for optical counterparts of the ICRF2 sources, as described in Section~\ref{dr7}. We found 1297 matching objects (233 galaxies and 1064 quasars). The identified objects were evenly distributed within the SDSS DR8 sky coverage. However, while investigating the SDSS--VLBI position offsets for quasars, we found an anomaly in the declination differences. 

The distribution of optical minus radio declination differences has two distinct peaks: one at $\mu_{\Delta\delta1}$$=$$2$~mas ($\sigma_{\Delta\delta1}$$=$$43$~mas), and another one at $\mu_{\Delta\delta2}$$=$$261$~mas ($\sigma_{\Delta\delta2}$$=$$54$~mas). The right ascension differences show no such obvious anomaly, with a normal distribution characterised by $\mu_{\Delta\alpha}$$=$$-9$~mas and $\sigma_{\Delta\alpha}$$=$$60$~mas. 
The sources contributing to the anomalous secondary peak are almost exclusively found above $\sim$$40\degree$ declination. This indicates a systematic error in the astrometric calibration of SDSS DR8.
Indeed, our independent result was confirmed and the reasons detailed by \citet{aihara2}, who reported an ``offset by roughly 250~mas in a northwest direction'' in a region that ``has irregular boundaries but in places extends as far south as declination $\delta$$\approx$$41\degree$''. The authors recommend to use DR7 for astrometric purposes where available, until the DR8 calibration is repaired.

\section{Conclusions}

SDSS DR7, although not an astrometric survey, could be used to compare optical and radio coordinates of 806 quasars, all with accurate positions determined with VLBI at the mas or sub-mas level. We confirmed that the SDSS coordinates of these objects agree in both right ascension and declination with $\sigma$$\approx$50~mas. We found 23 quasars for which the radio and optical brightness peaks differ by more than 155~mas (3$\sigma$). Before the accurate optical (Gaia) and radio (VLBI) reference frames are aligned in the future, it is essential to identify and filter out such peculiar sources from the link objects, and also study them from an astrophysical point of view. Here we proposed possible astrophysical explanations for the large non-coincidences: dual AGNs and gravitational lensing. These sources should be further investigated in detail to identify the cause(s). Finally, we found a systematic error in the astrometric calibration of the latest SDSS DR8, at declinations $\delta\ga40\degree$.

\begin{acknowledgements}
This work was supported by the Hungarian Scientific Research Fund (OTKA K72515). G.O. is grateful for the travel support received from the organisers of the GREAT-ESF workshop. The SDSS Web Site is \url{http://www.sdss.org}, the SDSS-III web site is \url{http://www.sdss3.org}, where the lists of their funding organisations and collaborating institutions can be found. 
\end{acknowledgements}

\bibliographystyle{aa}

\end{document}